\documentclass[aps,prl,twocolumn]{revtex4-1}

\usepackage{graphicx}
\newcommand{\vk}{\vec{k}}
\newcommand{\vx}{\vec{x}}
\newcommand{\vy}{\vec{y}}
\newcommand{\ud}{\mathrm{d}}
\newcommand{\ar}{\arrowvert}
\newcommand{\ra}{\rangle}
\newcommand{\la}{\langle}

\newcommand{\be}{\begin{equation}}
\newcommand{\ee}{\end{equation}}
\newcommand{\ba}{\begin{eqnarray}}
\newcommand{\ea}{\end{eqnarray}}

\begin{document}

\title{Cubic neutrons}
\date{\today}

\author{Felipe J. Llanes-Estrada}
\affiliation{On leave at Theor. Phys. Dept., Technische Universit\"at Muenchen, 85747 Garching, Germany.}
\author{Gaspar Moreno Navarro}
\affiliation{Departamento de F\'{\i}sica Te\'orica I, Univ. Complutense de Madrid, 28040 Madrid, Spain.}

\begin{abstract}
The neutron is largely spherical and incompressible in atomic nuclei. These two properties are however challenged in the extreme pressure environment of a neutron star. 
Our variational computation within the Cornell model of Coulomb gauge QCD shows that the neutron (and also the $\Delta^{3/2}$ baryon) can adopt cubic symmetry at an energy cost of about 150 $MeV$.\\
Balancing this with the free energy gained by tighter neutron packing, we expose the possible softening of the equation of state of neutron matter.
\end{abstract}


\maketitle


The atomic nucleus is a close-packed system analogous to a cold quantum liquid.  Its hadronic components are very incompressible, as revealed by  the nuclear radius scaling with the mass number as $r_A \simeq r_0 A^{-1/3}$, with unit of length $r_0\simeq 1.2\ fm$. 
Nucleons there are quite optimally packed, as the (charge) nucleon radius has been precisely determined~\cite{Hill:2010yb} in electron scattering, and is approximately $0.88\ fm$. If we (grossly) think of the nucleon as a solid sphere, $74\%$ of the nuclear volume is occupied, near the Keplerian limit~\cite{Hales:1992} reached by hcp or fcc lattices. 

In neutron stars, gravity-compressed nuclear matter finds extreme pressure environments that changes some of its properties. Recently~\cite{Demorest:2010bx} a two-solar mass neutron star has been discovered that takes nuclear physics to the limit (and we have recently proposed~\cite{bindG} that this is so much so as to constrain gravity for strong field intensity).
At high pressure and density in the core of the star many exotic phases of hadron  matter have been  proposed,  including a neutron superconductor/superfluid, various meson condensates, quarkyonic or Color-Flavor-Locked condensates, etc. Some of them have already been ruled out by the new, superheavy neutron star~\cite{Lattimer:2010uk}.
\\ 
At the crust one expects a thin atomic sheet~\cite{LlanesEstrada:2009na}
followed by neutron-rich nuclear matter, and finally almost pure neutron matter (with a small number of protons in $\beta$-equilibrium)~\cite{Zeldovich}.

One intermediate phase that has long been proposed~\cite{Canuto:1972} is a neutron crystal.  
In analogy with condensed matter systems (saliently $^3He$, a quantum liquid through $T=0$ that solidifies upon compression) neutron matter is assumed to locally adopt a periodic lattice-like arrangement to maximize density. \\
Our new observation is that when the environment looses its isotropy, 
the nucleon (largely spherical at rest) is subject to directional stresses and may deform to minimize the evacuated volume. 
Thus, for mass-energy densities $\varepsilon$ larger than $140\ MeV/fm^3\simeq 3 m_\pi^4$ neutrons can no more be considered pointlike objects subject to the rules of local, rotationally-invariant quantum field theory as elementary fields. Their composition and structure should start being taken into account, perhaps revisiting some of the hypothesis underlying the nuclear many body problem approach based on QFT methods (particularly, rotational invariance).

We report an estimate of the deformability of the neutron with the renowned Cornell Hamiltonian, 
\ba
\mathcal{H}(\vx) = \Psi^\dagger(\vx) (-i\vec{\alpha}\cdot\nabla + m
\beta) \Psi 
+ \\ \nonumber
\frac{1}{2} \int \ud^3 y \Psi^\dagger(\vx) T^a \Psi(\vx)
V(\vx-\vy) \Psi^\dagger(\vy) T^a \Psi(\vy)
\ea
a field theory model of Quantum Chromodynamics, tractable by many-body methods common in nuclear and condensed matter theory. The relativistic quark fields $\Psi$ interact via the color charge density $\Psi^\dagger(\vx) T^a \Psi(\vx)$. The model has been amply exploited to treat the heavy quarkonium spectrum~\cite{Eichten:1978tg,TorresRincon:2010fu}, as well as both light mesons and baryons~\cite{LlanesEstrada:1999uh,Bicudo:2009cr} with the Cornell potential $V(r)=\sigma r - \frac{4\alpha_s}{3r}$.

\begin{figure*}
\includegraphics[width=2in,angle=-90]{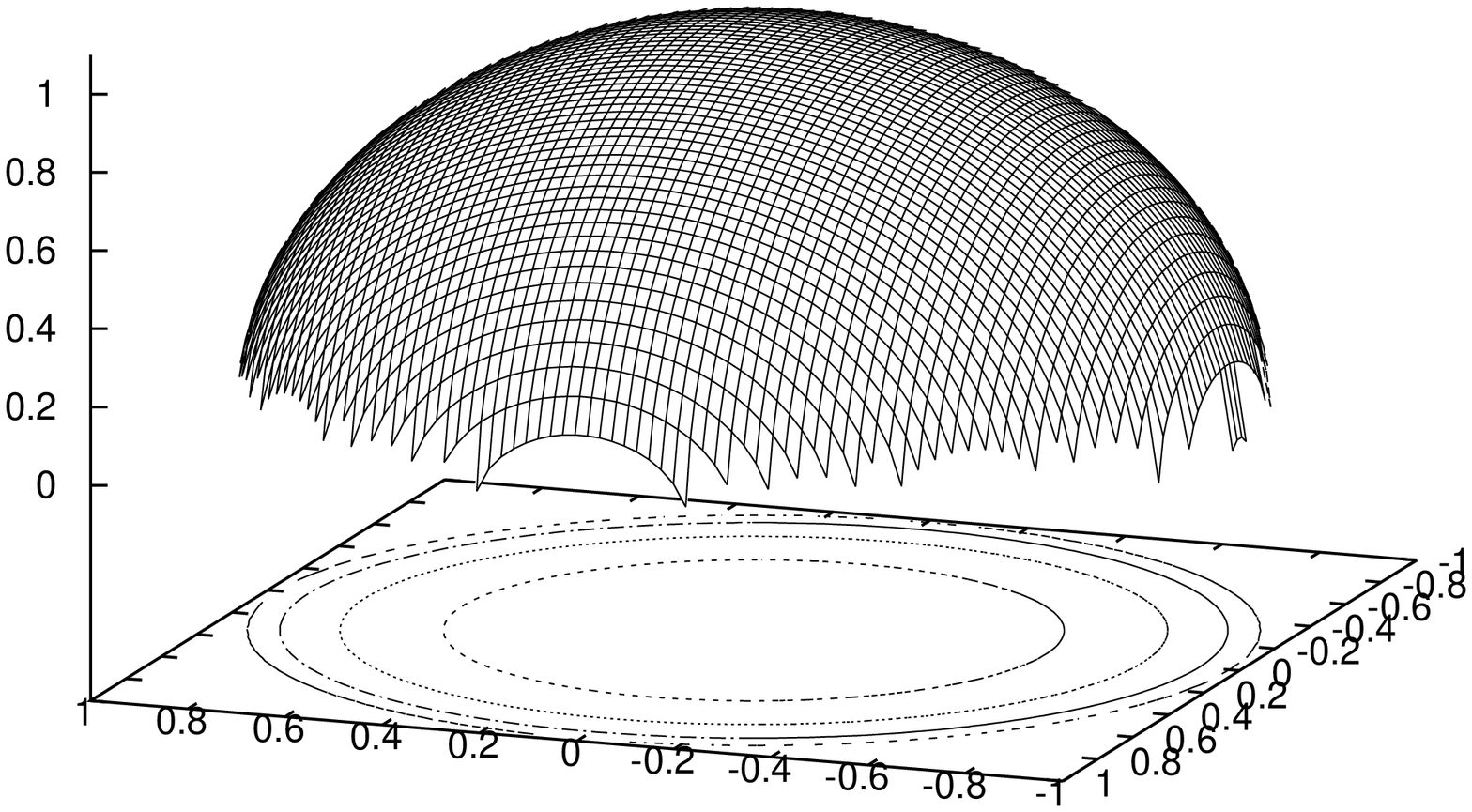}
\includegraphics[width=2in,angle=-90]{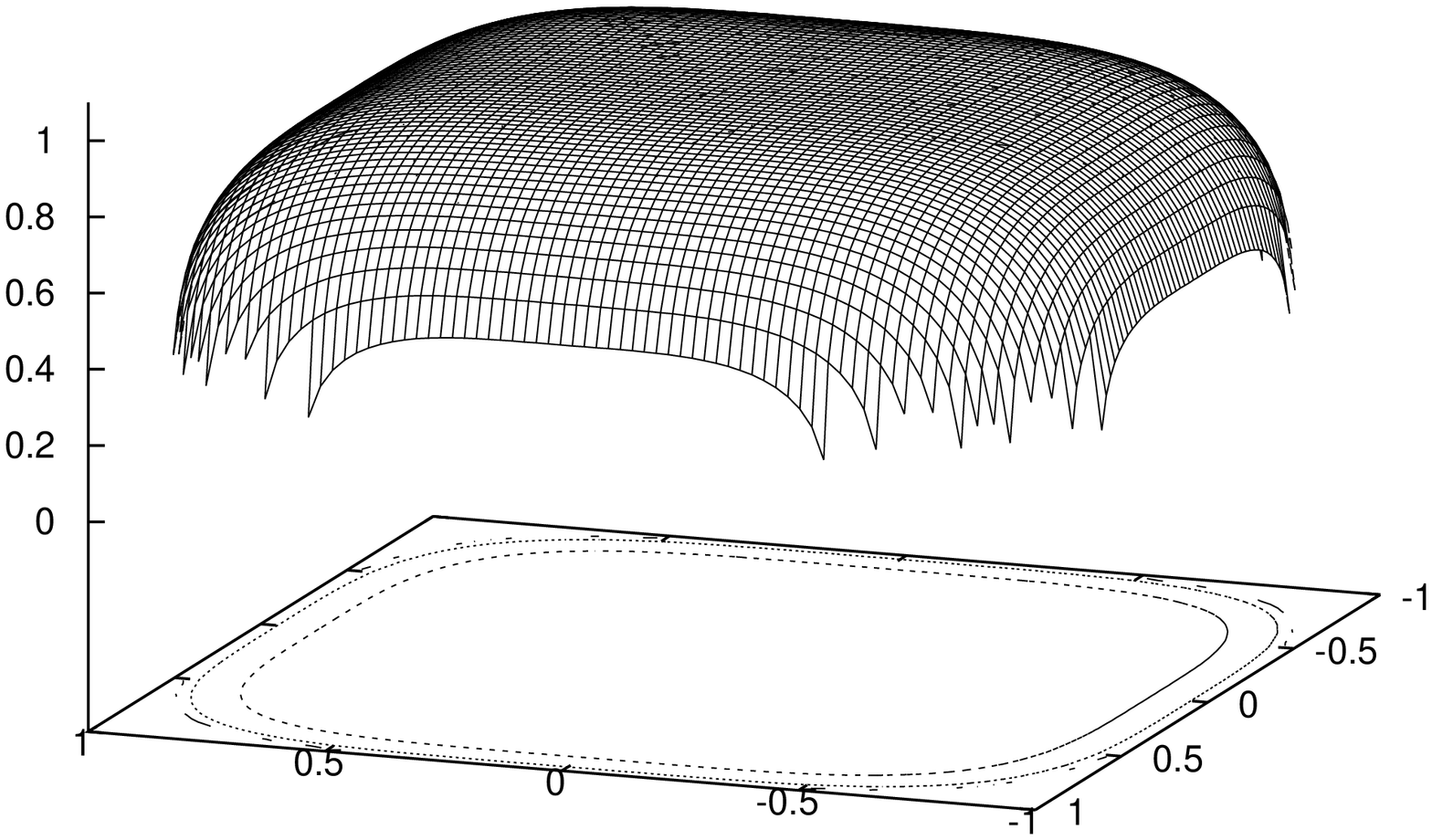}
\includegraphics[width=2in,angle=-90]{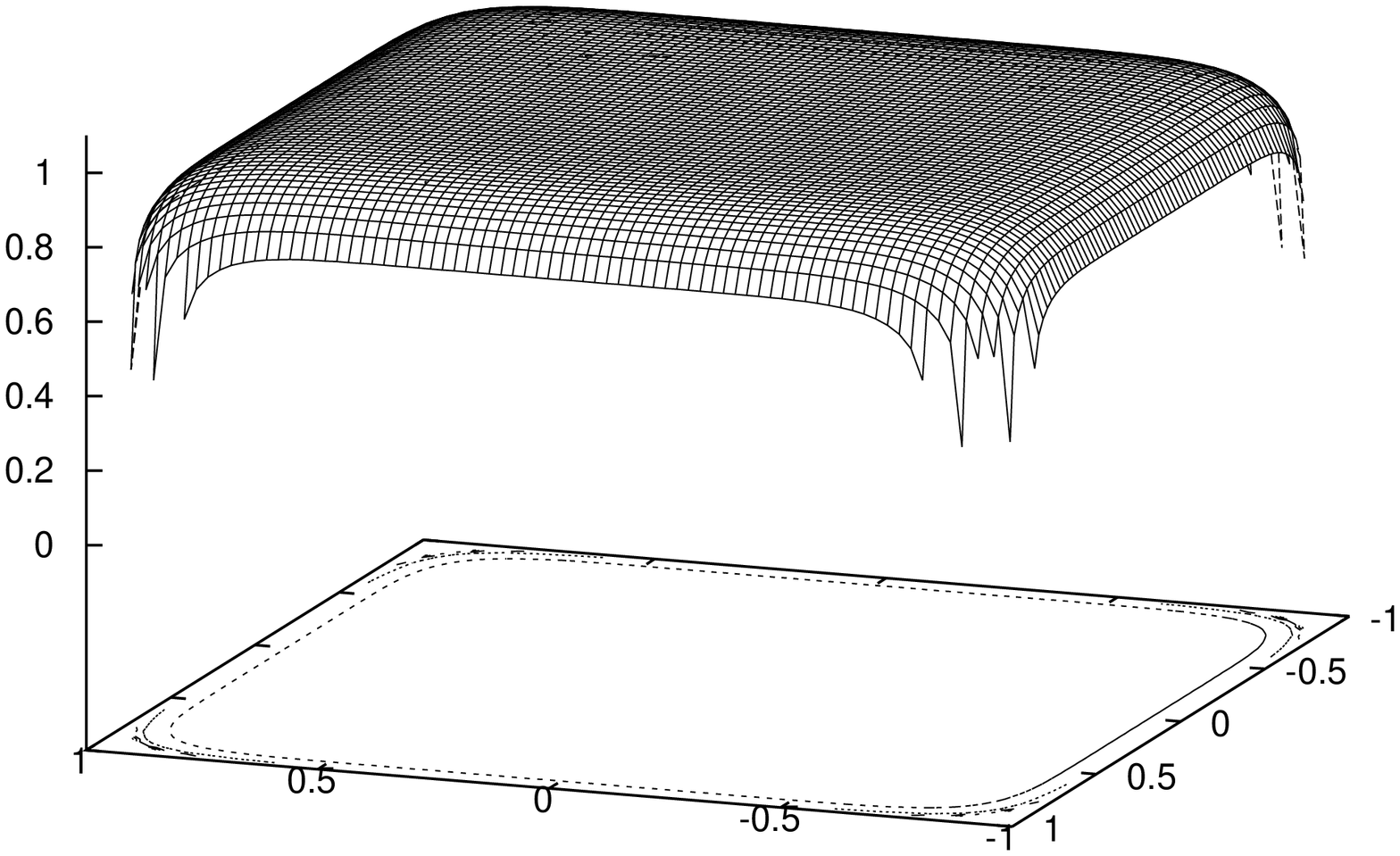}
\includegraphics[width=2in,angle=-90]{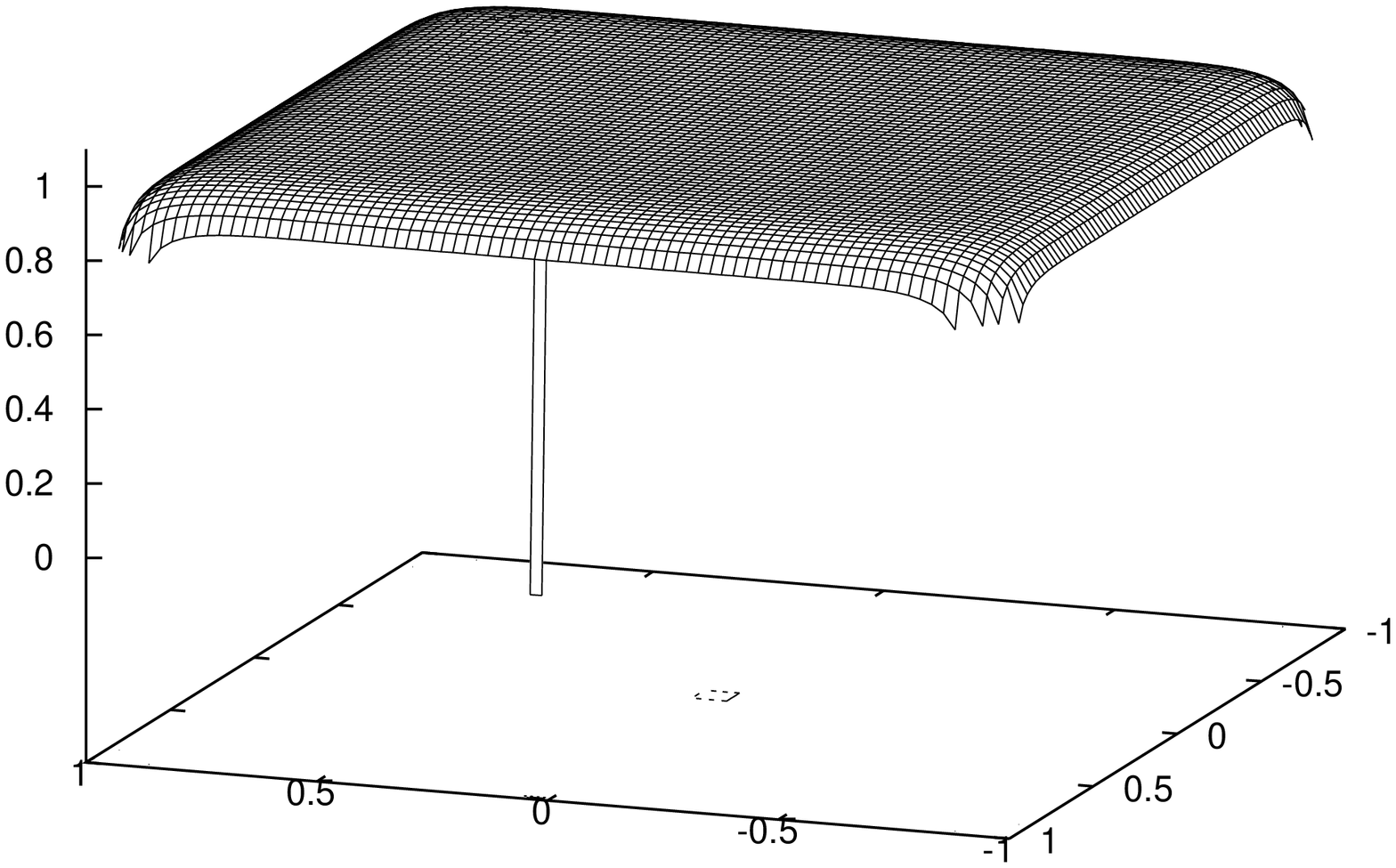}
\caption{Trial wavefunction that interpolates between sphere (for $N=2$), and cube (as $N\to\infty$) for $N=2,4,8,12$. \label{fig:cubosfera}}
\end{figure*}

The model ground state ($^3P_0$ quark-antiquark condensed vacuum) is treated in BCS approximation to generate constituent quarks from the current $m=5\ MeV$ quarks in the Hamiltonian, then
the second-quantized wavefunction appropriate for 3-quark baryons in terms of Bogoliubov-rotated quark creation operators $B^\dagger$ is
\be
\ar N \ra = \frac{\epsilon^{ijk}}{6}
F_N^{\lambda_1\lambda_2\lambda_3}(\vk_1,\vk_2,\vk_3)
B_{\vk_1\lambda_1i}^+ B_{\vk_2\lambda_2j}^+ B_{\vk_3\lambda_3k}^+ |
\Omega>\  .
\ee
To approach lowest-lying baryons by the Rayleigh-Ritz variational principle, we employ the separable  wavefunction (${\bf \rho}$, ${\bf \lambda}$ being 3-body Jacobi coordinates with Cartesian components ${\bf \rho}_x$, ${\bf \lambda}_z$, etc.)
\be \label{cubicwf}
F_N({\bf k}_\rho,{\bf k}_\lambda)  =  \psi_N\left( \frac{{\bf k}_\rho}{\alpha_{\rho}} \right)  \psi_N\left( \frac{{\bf k}_\lambda}{\alpha_{\lambda}} \right)
\ee
To choose a convenient variational wavefunction we observe that the graph of $(k_x^N+k_y^N+k_z^N)^{\frac{1}{N}}=1$, which is depicted in figure~\ref{fig:cubosfera} for even $N$, interpolates between spherical symmetry for $N=2$ and octahedral (the cube's) symmetry, reached as $N\to \infty$. Therefore one can take an appropriately normalized $\psi_N$ from a precise numerical solution of the two-body problem and use $\psi_N\left((k_x^N+k_y^N+k_z^N)^{\frac{1}{N}}\right)$ to study the three-quark system transiting between both symmetry groups.

The Hamiltonian expectation value $\la N\ar H\ar N\ra$ is a function of the two variational parameters $\alpha_\rho$, $\alpha_\lambda$, and is minimized respect to them to find the best approximation to the neutron mass within this family of functions for $N=2$. To compute the matrix elements we perform all relativistic spinor sums numerically, and the three-body nine-dimensional integral by Monte Carlo methods~\cite{Hahn:2004fe}. The model approximation to the nucleon mass is 980 $MeV$ (with a Monte Carlo error of $40\ MeV$), the physical mass being $940\ MeV$. Should higher precision
be necessary one should resort to lattice gauge theory~\cite{Durr:2008zz}. This is not so for our purposes and we proceed to change the wavefunction symmetry, repeating the minimization for varying $N$.

\begin{figure}
\includegraphics[width=3.in]{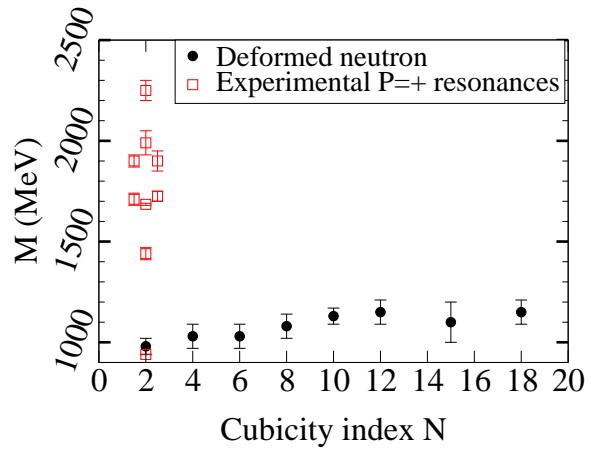}
\caption{Open squares: experimental neutron mass and known positive parity resonance excitations~\cite{Nakamura:2010zzi}. Filled circles: variational minimum for the nucleon mass with the wavefunction in Eq.~(\ref{cubicwf}). At a cost of 50 $MeV$ the neutron already looses spherical symmetry, adopting an almost totally cubic shape for $150\ MeV$.
\label{fig:cubicneutron}}
\end{figure}
We find that the neutron mass increases by about $150\ MeV$ between $N=2$ and $N=18$ (the last already very close to a cubic neutron), as depicted in figure~\ref{fig:cubicneutron}. 

\begin{figure}
\includegraphics[width=3.in]{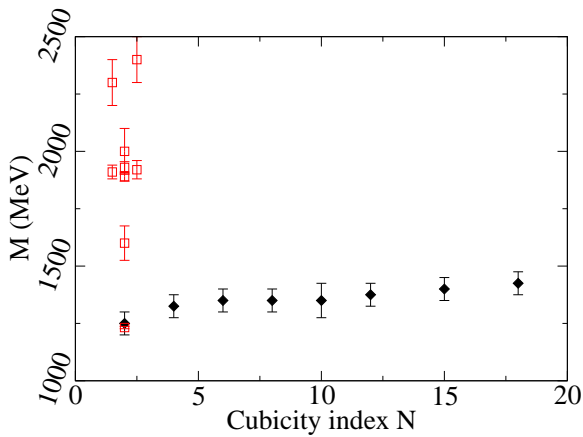}
\caption{Open squares: experimental $\Delta$-baryon mass and known positive parity resonance excitations~\cite{Nakamura:2010zzi}. Filled circles: variational minimum for the $\Delta$ mass with the wavefunction in Eq.~(\ref{cubicwf}). The excitation cost to abandon spherical symmetry is very similar to that reported in figure~\ref{fig:cubicneutron} for the neutron.
\label{fig:cubicDelta}}
\end{figure}

As a check of the spin-independence of our result, we have repeated the calculation for the $\Delta(1232)$ baryon, where all three quark spins are parallel, and obtained an equal excitation energy. The similar outcome is reported in figure~\ref{fig:cubicDelta}.

We can offer the following heuristic argument supporting these numerical findings. If one takes a high $N$ (close to cubic) variational wavefunction, and performs a spherical harmonic analysis,  75\% of the probability resides in the $Y_{00}$ spherically symmetric state, and an additional 24\% in one of the d-wave $Y_{2m}$ (mostly $m=0$) states. 
Weighing the experimental masses (several positive-parity excitations of the nucleon are also shown in figure \ref{fig:cubicneutron}) 
of ground $J^P=\frac{1}{2}^+$ (940 $MeV$) and d-wave excited $J^P=\frac{5}{2}^+$ (1680 $MeV$) nucleons in the proportion 3:1 yields 1125 $MeV$. This entails a cost of about 180 $MeV$ to change the neutron symmetry from spherical to cubic. Thus, from a knowledge of the experimental spectrum and symmetry considerations alone one can already obtain the approximate answer. 
The same reasoning applies to the  150 $MeV$ excitation energy of the $\Delta$, that can likewise be interpreted since the relevant d-wave resonance energy is tagged by the $\Delta^{5/2 +}(1900)$ and the $\Delta^{7/2 +}(1950)$.

\begin{figure}
\includegraphics[width=3.in]{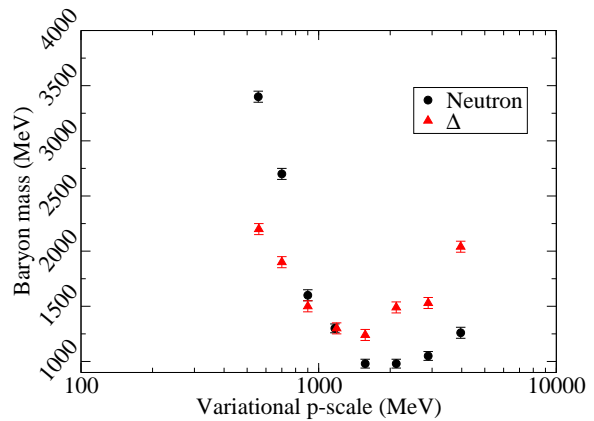}
\caption{Mass of nucleon and $\Delta$-baryon as function of only one variational parameter $\alpha_\rho=\alpha_\lambda$. In addition to a minimum corresponding to the best Rayleigh-Ritz representation of the particles within the function family, the energy cost of compression or dilatation of the wave function (at constant shape) can be appreciated.
\label{fig:compress}}
\end{figure}

Although deformability at constant hadron volume, and not compressibility of the hadron, was the goal of our investigation, we can provide a feeling for the latter by setting both Jacobi scales equal $\alpha_\rho=\alpha_\lambda$ and studying the mass as function as this variational parameter alone. We refer the reader to figure~\ref{fig:compress}. A variational minimum is clearly seen for both nucleon and $\Delta$, corresponding to their best Rayleigh-Ritz representation. Compressing or dilating the neutron can be faked by varying this scale, and the cost of doing so is evident in the figure, and of the order of magnitude that one naively expects. The neutron seems to be somewhat stiffer to momentum-space compression (position-space dilatation), while both particles react similarly to momentum-space dilatation (position-space compression).

The excitation energy that we have found is almost equal to that necessary to produce hyperons, since the lightest strange particle, $\Lambda$, has a mass of about 1115 $MeV$. Thus, if the nucleon's Fermi momentum is high enough to relax some of the $d$ quarks to $s$ quarks, it is also high enough to deform the nucleon, and both phenomena appear simultaneously. Their interplay is beyond our current undertaking.

\begin{figure}
\vspace{0.3cm}
\includegraphics[width=2.8in]{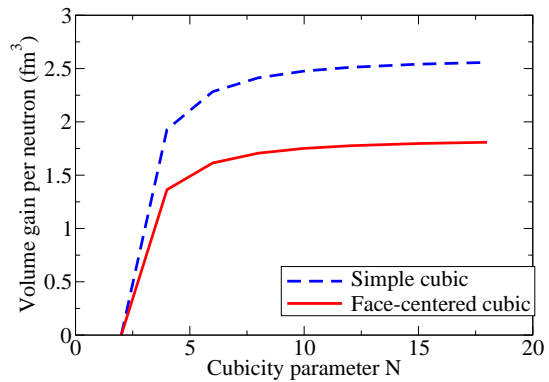}
\caption{Interstitial volume gain $\Delta V$ by deformation of the neutron wavefunction \emph{at constant neutron volume} for two cubic lattices\label{fig:volume}. The gain is naturally smaller for the more packing-efficient face-centered-cubic lattice, but still significant.
}
\end{figure}

\begin{figure}
\includegraphics[width=3.in]{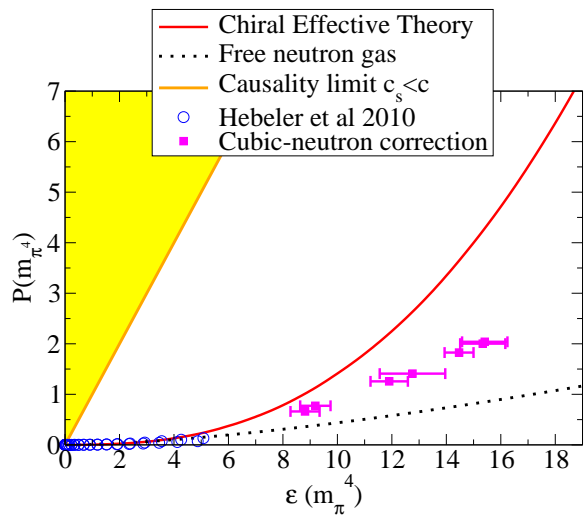}
\caption{Effect on the state of the art equation of state\cite{Lacour:2009ej,Hebeler:2010jx} of neutron matter of the deformability of the neutron.
The entire interval shown is relevant for neutron star structure. Nucleon deformability produces a softening of the equation of state additional to other known sources\label{stateeq}.}
\end{figure}

At high enough pressure, the $150\ MeV$ energetic cost to deform the neutron can be provided by the Helmholtz free energy $P\Delta V$ obtained upon reducing the interstitial volume between neutrons thanks to the cubic shape. The maximum gain as $N\to \infty$ in an fcc lattice, with packing fraction 0.74, is 26\% of the initial volume occupied by a spherical neutron.
Estimating this volume for arbitrary $N$ through simple three-dimensional integration (see figure~\ref{fig:volume}) 
we can, at fixed pressure, obtain the consequent increase in density. 
We plot in figure \ref{stateeq} the effect on the equation of state of neutron matter of the neutron deformability that we have found. 
The equation of state becomes softer (smaller pressure at given density) at densities relevant for neutron star structure. 


To conclude, let us observe that the bulk properties of neutron matter in the star will be affected by the neutrons adopting cubic symmetry (and patches of neutron matter becoming crystalline), particularly the Young modulus, the moment of inertia, etc. and this will surely provide detailed tests  for future work.

\begin{acknowledgments}
FJLE thanks a Caja Madrid fellowship for advanced study and the hospitality of the theory group at TU-Munich and the Exzellenzcluster „Origin and Structure of the Universe“.  This work has been supported by grants
 227431-HadronPhysics2 (EU), Consolider-CSD2007-00042, AIC10-D-000582,       FPA2008-00592, FIS2008-01323,  and UCM-BSCH GR58/08 910309 (Spain).
\end{acknowledgments}

\bibliography{Cubicneutrons.bib}

\end{document}